\documentclass[twocolumn,aps,superscriptaddress,showpacs,floatfix]{revtex4-1}
\usepackage{graphicx}
\usepackage{dcolumn}
\usepackage{bm}
\usepackage[colorlinks,linkcolor=blue,anchorcolor=blue,citecolor=blue,urlcolor=blue]{hyperref}

\begin{document}

\preprint{APS/123-QED}

\title{Systematic study of unfavored $\mathcal{\alpha}$ decay half-lives of closed shell nuclei related to ground and isomeric states}

\author{Jun-Gang Deng}
\affiliation{School of Nuclear Science and Technology, University of South China, 421001 Hengyang, People's Republic of China}
\author{Jie-Cheng Zhao}
\affiliation{School of Nuclear Science and Technology, University of South China, 421001 Hengyang, People's Republic of China}
\author{Dong Xiang}
\email{xiangdong007@163.com }
\affiliation{School of Nuclear Science and Technology, University of South China, 421001 Hengyang, People's Republic of China}
\author{Xiao-Hua Li}
\email{lixiaohuaphysics@126.com }
\affiliation{School of Nuclear Science and Technology, University of South China, 421001 Hengyang, People's Republic of China}
\affiliation{Cooperative Innovation Center for Nuclear Fuel Cycle Technology $\&$ Equipment, University of South China, 421001 Hengyang, People's Republic of China}
\affiliation{Key Laboratory of Low Dimensional Quantum Structures and Quantum Control, Hunan Normal University, 410081 Changsha, People's Republic of China}

\begin{abstract}
In present work, the unfavored $\mathcal{\alpha}$ decay half-lives and $\mathcal{\alpha}$ preformation probabilities of closed shell nuclei related to ground and isomeric states around $Z=82$, $N=82$ and 126 shell closures are investigated by adopting two-potential approach from the perspective of valence nucleon (hole) and isospin asymmetry of the parent nucleus. The results indicate that $\mathcal{\alpha}$ preformation probability is linear dependence on $N_p N_n$ or $N_p N_n I$, the same as the case of favored $\mathcal{\alpha}$ decay in our previous work [X.-D. Sun \textit{et al.}, Phys. Rev. C 94, 024338 (2016)]. $N_p$, $N_n$, and $I$ represent the number of valence protons (holes), the number of valence neutrons (holes), and the isospin of the parent nucleus, respectively. Fitting the $\mathrm{\alpha}$ preformation probabilities data extracted from the differences between experimental data and calculated half-lives without a shell correction, we give two linear formulas of the $\mathcal{\alpha}$ preformation probabilities and the values of corresponding parameters. Based on the formulas and corresponding parameters, we calculate the $\mathrm{\alpha}$ decay half-lives for those nuclei. The calculated results can well reproduce the experimental data.

\end{abstract}

\maketitle

\section{Introduction}
$\mathcal{\alpha}$ decay was defined in 1899 by Rutherford, and the quantum tunnel theory was independently put forward to estimate the probability of an $\mathcal{\alpha}$ particle tunneling through the Coulomb barrier in 1928 by Gurney and Condon \cite{Gur28} and Gamow \cite{Gamow1928}. Since then, $\mathcal{\alpha}$ decay, as one of the most important tools to study unstable nuclei, neutron-deficient nuclei and superheavy nuclei, has been a hot area of research in nuclear physics \cite{PhysRevLett.112.172501,PhysRevLett.112.092501,PhysRevC.92.051301,Ni2015108,PhysRevC.84.064608, PhysRevC.85.044608}. Theoretically, $\mathcal{\alpha}$ decay shares the similar theory of barrier penetration with different kinds of charged particles' radioactivity, for instance, heavy ion emission, single proton emission, spontaneous fission \cite{0954-3899-17-6-001,PhysRevC.80.024310,PhysRevC.83.014601,PhysRevC.79.054330,SANTHOSH201249,0954-3899-42-8-085101}, and so on. Experimentally, $\mathcal{\alpha}$ decay is the main decay mode for most of the new synthesis of superheavy nuclei and sometimes it is the unique. Meanwhile, for some very unstable new synthetic nuclei, $\mathcal{\alpha}$ decay is the effective way to determine their identities (the number of protons and neutrons).

Usually, $\mathcal{\alpha}$ decay is described as a process of a preformed $\mathcal{\alpha}$ particle tunneling through the potential barrier between $\mathcal{\alpha}$ cluster and the daughter nucleus, and the preformed probabilities of $\mathcal{\alpha}$ cluster are different for various nuclei. Therefore, the calculated $\mathcal{\alpha}$ decay constant should be multiplied by a preformation factor of $\mathcal{\alpha}$ particle. Nevertheless, we rarely know formation and movement of $\mathcal{\alpha}$ particle inside the parent nuclei, as a result of the complicated structure of the quantum many-body systems. Therefore, there are a few works \cite{QI201677,PhysRevC.77.054318,PhysRevC.92.044302, PhysRevC.95.061306} studying $\mathcal{\alpha}$ preformation probabilities from the viewpoint of microscopic theory. Phenomenologically, $\mathcal{\alpha}$ preformation probabilities are obtained by the ratios of theoretical calculations without considering the preformation factors to experimental half-lives \cite{PhysRevC.84.027303,Qian2013,PhysRevC.80.064325, PhysRevC.73.031301}. Recent research has shown that, the pairing effect, the shell effect, and different spin-parity states of daughter and parent nucleus are the major factors determining $\mathcal{\alpha}$ preformation probabilities \cite{PhysRevC.92.044302}. Seif \emph{et al.} have proposed that the $\mathcal{\alpha}$ preformation probability, considering isospin, shell effect and valence proton-neutron interaction, is proportional to $ N_p N_n$ for even-even nuclei around proton $Z=82$, neutron $N=82$ and 126 shell closures \cite{PhysRevC.84.064608}, where $ N_p$, $N_n$ denote valence protons (holes) and valence neutrons (holes) of parent nucleus, respectively. Furthermore, in our previous work \cite{PhysRevC.94.024338}, the $\mathcal{\alpha}$ preformation probabilities of odd-$A$ and doubly-odd nuclei favored $\mathcal{\alpha}$ decay also satisfy this relationship. Therefore, it is interesting to validate whether this linear relationship still exist in unfavored $\mathcal{\alpha}$ decay of closed shell nuclei. In this work, we investigate the $\mathcal{\alpha}$ decay half-lives and $\mathcal{\alpha}$ preformation probabilities for unfavored $\mathcal{\alpha}$ decay of closed shell nuclei around $Z=82$, $N=82$ and 126 closed shells, respectively. Our results indicate that in unfavored $\mathcal{\alpha}$ decay of closed shell nuclei, the $\mathcal{\alpha}$ preformation probabilities are still linear related with $N_p N_n$, i.e. valence proton-neutron interaction plays an important role on $\mathcal{\alpha}$ preformation probabilities. The calculated results can well reproduce the experimental data from NUBASE2012 \cite{1674-1137-36-12-001}.

This article is organized as follows. In next section, the theoretical framework of the $\mathcal{\alpha}$ decay half-life is briefly presented. The detailed calculations and discussions are given in Sec. \ref{section 3}. In this section, we investigate the $\mathcal{\alpha}$  preformation probabilities from the viewpoint of the valence proton-neutron interaction and isospin effect, respectively. Sec. \ref{section 4} is a brief summary.

\section{THEORETICAL FRAMEWORK}
$\mathcal{\alpha}$ decay half-life $T_{1/2}$, as an important indicator of nuclear stability, can be calculated by the $\mathcal{\alpha}$ decay width $\Gamma$ and written as
\begin{equation}
\
T_{1/2}=\frac{{\hbar}ln2}{\Gamma}
.\label{subeq:1}
\end{equation}
In general, $\mathcal{\alpha}$ decay can be approximatively treated as a stationary state problem, on account of the $\Gamma$ is much smaller than the $\mathcal{\alpha}$ decay energy $Q_\alpha$. Hence, we can adopt two-potential approach (TPA), which has been successfully applied to deal with metastable states, to calculate $\mathcal{\alpha}$ decay half-life \cite{PhysRevLett.59.262,PhysRevC.85.027306,QIAN201182,QIAN20111, PhysRevC.95.014319, PhysRevC.95.044303}. In the framework of TPA, the $\mathcal{\alpha}$ decay width is calculated as below
\begin{equation}
\
\Gamma=P_{\alpha}\frac{{\hbar}^2FP}{4\mu}
,\label{subeq:1}
\end{equation}
where $\mu$ is the reduced mass of daughter nucleus and $\mathcal{\alpha}$ particle, and $\mu=\frac{{m_d}{m_{\alpha}}}{{m_d}+{m_{\alpha}}}$ with ${m_d}$ and ${m_{\alpha}}$ being mass of daughter nucleus and $\mathcal{\alpha}$ particle, respectively.

$P$ is the semiclassical Wentzel-Kramers-Brillouin (WKB) barrier penetrate probability, namely, Gamow factor. It can be expressed as
\begin{equation}
\
P=\exp(-2{\int_{r_2}^{r_3} k(r) dr})
,\label{subeq:1}
\end{equation}
where $\mathrm{ k(r)=\sqrt{\frac{2\mu}{{\hbar}^2}|Q_{\alpha}-V(r)|}}$ is the wave number of the $\mathcal{\alpha}$ particle , and $\mathrm{r}$ is the center of mass distance between the daughter nucleus and the preformed $\mathcal{\alpha}$ particle. $V(r)$ is the $\mathcal{\alpha}$-core potential.

The normalized factor $F$, denoting the assault frequency of $\mathcal{\alpha}$ particle, is approximatively calculated by
\begin{equation}
\
F{\int_{r_1}^{r_2} \frac{1}{2k(r)} dr}=1,
\label{subeq:1}
\end{equation}
where $r_1$, $r_2$ and the above $r_3$ are the classical turning points, they can be obtained through solving $V (r_1) = V (r_2) = V (r_3) =Q_\alpha$.

$P_{\alpha}$ means the $\mathcal{\alpha}$ preformation probability, recent researches indicate that it rapidly decline in near closed shell and mildly change in an open shell region \cite{PhysRevC.84.027303,Qian2013,PhysRevC.80.064325}. $P_{\alpha}$ increases with the increase of valence nucleons up to next closed shell, and decreases with the decrease of valence holes. Usually, the value of $P_{\alpha}$ is defined as $P_{\alpha}=P_0\frac{T_{1/2}^{\text{calc}}}{T_{1/2}^{\text{expt}}}$, where $T_{1/2}^{\text{expt}}$ denotes experimental half-life, $T_{1/2}^{\text{calc}}$ represents the calculated $\mathcal{\alpha}$ decay half-life based on an assumption that $\mathcal{\alpha}$ preformation probability is a different constant for different kinds of nuclei. In accordance with the calculations by adopting the density-dependent cluster model (DDCM) \cite{XU2005303}, $P_0$ is 0.43 for even-even nuclei, 0.35 for odd-$\mathrm{A}$ nuclei, and 0.18 for doubly-odd nuclei.

The total interaction potential $V(r)$, which is composed of nuclear potential $V_N(r)$, Coulomb potential $V_C(r)$, and centrifugal potential $V_l(r)$, can be written as
\begin{equation}
\
V(r)=V_N(r)+V_C(r)+V_l(r)
.\label{subeq:1}
\end {equation}
In this work, we choose a cosh parametrized form for the nuclear potential $V_N(r)$, obtained by analyzing experimental data of $\mathcal{\alpha}$ decay \cite{PhysRevLett.65.2975}, which can be expressed as
\begin{equation}
\
V_N(r)=-V_0\frac{1+\mathrm{cosh}(R/a_0)}{\mathrm{cosh}(r/a_0)+\mathrm{cosh}(R/a_0)},
\label{subeq:1}
\end {equation}
where $V_0$ and $a_0$ mean the depth and diffuseness of the nuclear potential, respectively. In our previous work, we have obtained a set of parameters by analyzing the 164 even-even nuclei experimental $\mathcal{\alpha}$ decay half-lives data, which is $a_0$ =0.5958 fm and $V_0=192.42+31.059\frac{N_d-Z_d}{A_d}$ MeV \cite{PhysRevC.93.034316} with the $N_d$, $Z_d$, and $A_d$ being the number of neutrons, protons and mass number of the daughter nucleus, respectively. In this work, we also adopt these parameters to calculate the nuclear potential.

$V_C(r)$, the Coulomb potential, is regarded as the potential of a uniformly charged sphere with sharp radius $R$, which can be expressed as
\begin{equation}
\
V_C(r)=\left\{\begin{array}{ll}

\frac{Z_dZ_{\alpha}e^2}{2R}[3-(\frac{r}{R})^2],&\text{{r}\textless{R}},\\

\frac{Z_dZ_{\alpha}e^2}{r},&\text{{r}\textgreater{R}},

\end{array}\right.
\label{subeq:1}
\end {equation}
where $Z_\alpha$=2 is the number of protons for preformed $\mathcal{\alpha}$ particle. The centrifugal potential $V_l(r)$ can be estimated by
\begin{equation}
\
V_l(r)=\frac{{\hbar}^2(l+1/2)^2}{2{\mu}r^2}
,\label{subeq:1}
\end {equation}
where $l$ is the orbital angular momentum taken away by the $\mathcal{\alpha}$ particle. $l=0$ for the favored $\mathcal{\alpha}$ decays, while $l{\ne}0$ for the unfavored decays. Adopting the Langer modified centrifugal barrier, for one-dimensional problems, $l(l+1){\to}(l+1/2)^2$ is a necessary corrections \cite{1995JMP....36.5431M}. According to the conservation law of angular momentum \cite{PhysRevC.79.054614}, the minimum angular momentum $l_{\text{min}}$ taken away by the $\mathcal{\alpha}$ particle can be obtained by
\begin{equation}
\
l_{\text{min}}=\left\{\begin{array}{llll}

{\Delta}_j,&\text{for even${\Delta}_j$ and ${\pi}_p$= ${\pi}_d$},\\

{\Delta}_j+1,&\text{for even${\Delta}_j$ and ${\pi}_p$$\ne$${\pi}_d$},\\

{\Delta}_j,&\text{for odd${\Delta}_j$ and  ${\pi}_p$$\ne$${\pi}_d$},\\

{\Delta}_j+1,&\text{for odd${\Delta}_j$ and  ${\pi}_p$= ${\pi}_d$},

\end{array}\right.
\label{subeq:1}
\end {equation}
where ${\Delta}_j= |j_p-j_d|$, $j_p$, $\pi_p$, $j_d$, $\pi_d$ are spin and parity values of the parent and daughter nuclei, respectively.

The sharp radius $R$ is calculated by
\begin{equation}
\
R=1.28A^{1/3}-0.76+0.8A^{-1/3}
.\label{subeq:1}
\end {equation}
The above empirical formula is derived from the nuclear droplet model and proximity energy \cite{0954-3899-26-8-305}. $A$ is the mass number of parent nucleus.

\section{RESULTS AND DISCUSSIONS}
\label{section 3}

\begin{figure}[b]
\includegraphics[width=8.5cm]{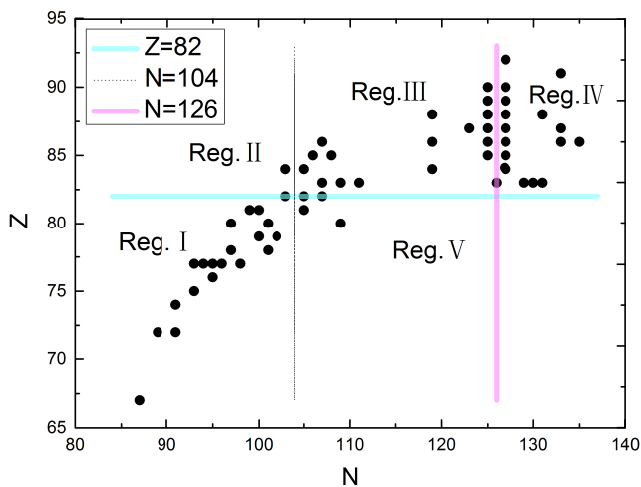}
\caption{(color online) Nuclide chart is divided into five regions. The cyan and magenta lines denote the $Z=82$, $N=126$ nuclear shell closures, respectively. The black dotted line represents $N=104$.}
\label{fig 1}
\end{figure}

The aim of present work is to study the effect of valence nucleons (holes) and isospin on $\mathcal{\alpha}$ preformation probabilities and half-lives of unfavored $\mathcal{\alpha}$ decay belong to nuclei around the $Z=82$, $N=82$ and 126 shell closures. For the odd-$A$ and doubly-odd nuclei, excitation of single nucleon causes the high-spin isomers. Our previous works \cite{1674-1137-41-1-014102,PhysRevC.94.024338} indicate that both ground and isomeric states can be treated in a unified way for $\mathcal{\alpha}$ decay parent and daughter nuclei.

Many researches show that the more smaller valence nucleons (holes) nuclei have, the more smaller $\mathcal{\alpha}$ preformation probabilities should be \cite{PhysRevC.84.027303,Qian2013,PhysRevC.80.064325}. In addition, valence proton-neutron interaction approximately remain unchanged in the same shell region \cite{PhysRevLett.85.720}. Recently, it is found that $\mathcal{\alpha}$ preformation probabilities are linear with product of valence protons (holes) and valence neutrons (holes) for even-even nuclei around $Z=82$, $N=82$ and 126 closed shells. Furthermore, Self $\textit{et al.}$ propose that isospin asymmetry also makes an important contribution to $\mathcal{\alpha}$ preformation probabilities \cite{PhysRevC.84.064608}. Moreover, our previous work \cite{PhysRevC.94.024338} indicates that the linear relationship between $P_{\alpha}$ and $N_p N_n$ still exist for the favored $\mathcal{\alpha}$ decay of odd-$A$ and doubly-odd nuclei in the same regions. However, for unfavored $\mathcal{\alpha}$ decay, it is more difficult than its counterparts because of the unpaired nucleon and different spin-parity state of parent and daughter nucleus. So it is necessary to study whether $\alpha$ preformation probabilities of unfavored $\mathcal{\alpha}$ decay of closed shell nuclei still exist the liner relationship with $N_p N_n$.

For more intuitively study, we plot a nuclide distribution map in Fig. \ref{fig 1}, and the area are marked as Region I to V in accordance with valence nucleons (holes), respectively. In this paper, we concentrate on Regions I, III, and IV, on account of a little number of nuclei with $\mathcal{\alpha}$ radioactivity
 in Regions II and V. For purpose of a deeper insight into relationship between $\mathcal{\alpha}$ preformation probabilities and $N_p N_n$, we study from standpoint of nuclear shell and isospin asymmetry, respectively within Eq. (\ref{subeq:11}) and (\ref{subeq:12}).
\begin{equation}
\
P_{\alpha}=a\frac{N_p N_n}{Z_0+N_0}+b
,\label{subeq:11}
\end {equation}

\begin{equation}
\
P_{\alpha}=cN_p N_nI+d
,\label{subeq:12}
\end {equation}
where $Z_0$, $N_0$ are adjacent magic number of proton and neutron, respectively. $a$, $b$, $c$ and $d$ are adjustable parameters extracted from the fittings of $P_{\alpha}$ of Table \ref{table 1}-\ref{table 3} and listed in the Table \ref{table 4} (the upper part for the case of odd-$A$ and bottom half for doubly-odd nuclei). Based on Eq. (\ref{subeq:11}), (\ref{subeq:12}) and corresponding parameters, we calculate the $\mathcal{\alpha}$ decay half-lives and express as $T^{\text{calc2}}_{1/2}$, $T^{\text{calc3}}_{1/2}$, respectively, which are listed in the last two columns of Table \ref{table 1}-\ref{table 3}. The first seven columns of Table \ref{table 1}-\ref{table 3} are $\mathcal{\alpha}$ transition, $\mathcal{\alpha}$ decay energy $Q_{\alpha}$, spin-parity transformation, the minimum orbital angular momentum $l_{\text{min}}$ taken away by $\mathcal{\alpha}$ particle, the experimental half-life $T^{\text{expt}}_{1/2}$, calculated half-life $T^{\text{calc1}}_{1/2}$ by TPA with $P_{\alpha} =P_0$, extracted $\mathcal{\alpha}$ preformation probability $P_{\alpha}$, respectively.

\begin{figure}[b]
\includegraphics[width=8.5cm]{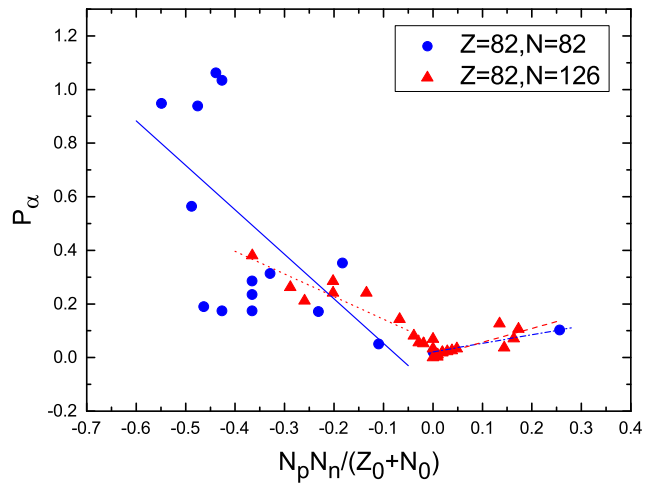}
\caption{(color online) The linear relationship between $\mathcal{\alpha}$ preformation probabilities and $\frac{N_p N_n}{Z_0+N_0}$. $N_p$, $N_n$ represent valence protons (holes) and neutrons (holes) of parent nucleus, respectively. $Z_0$, $N_0$ mean the magic numbers of proton and neutron, respectively. The blue solid and dash dot line denote the fittings of nuclei in Region I and, II, respectively. The red dot and dash line represent the fittings of nuclei in Region III, IV, respectively. }
\label{fig 2}
\end{figure}

\begin{table*}
\caption{Calculations of unfavored $\mathcal{\alpha}$ decay half-lives and the $\mathcal{\alpha}$ preformation probabilities of odd-$A$ nuclei in Region I around $Z=82$, $N=82$ closed shell. Elements with upper suffixes `m', `n' and `p' indicate assignments to excited isomeric states (defined as higher states with half-lives greater than 100 ns). Suffixes `p' also indicate non-isomeric levels, but used in the AME2012 \cite{1674-1137-36-12-003, 1674-1137-36-12-002}. `()' means uncertain spin and/or parity. `\#' means values estimated from trends in neighboring nuclides with the same $Z$ and $N$ parities.}
\label{table 1}
\begin{ruledtabular}
\begin{tabular}{ccccccccc}

{$\mathcal{\alpha}$ transition} & $Q_{\alpha}$ (MeV) &  ${j^{\pi}_{p}}\to{j^{\pi}_{d}}$ &$l_{\text{min}}$ & $T^{\text{expt}}_{1/2}$ (s) & ${T_{1/2}^{\text{calc1}}}$ (s)&${P_{\alpha}}$& ${T_{1/2}^{\text{calc2}}}$  (s)& ${T_{1/2}^{\text{calc3}}}$ (s)\\
 \hline
{}&{}&{}&{}&{}&{}&{}&{}&{}\\

$^{161}$Hf$\to^{157}$Yb$$&4.686&$ {3/2-\#}\to{7/2-}$ &2 & $1.82\times10^{4} $& $5.38\times10^{4}$ &1.034&$3.10\times10^{4}$&$2.80\times10^{4}$ \\
$^{163}$Hf$\to^{159}$Yb$$&4.150&$ {3/2-\#}\to{5/2(-)}$ &2 & $4.00\times10^{7}$ &$1.08\times10^{8}$ & 0.948 &$9.12\times10^{7}$&$8.61\times10^{7}$\\
$^{165}$W$\to^{161}$Hf$$&5.029&$ {(5/2-)}\to{3/2-\#}$ &2 & $2.55\times10^{3}$ & $7.74\times10^{3}$ & 1.062& $4.48\times10^{3}$& $3.94\times10^{3}$\\
$^{171}$Os$\to^{167}$W$$&5.371&$ {(5/2-)}\to{3/2-\#}$ &2 &$4.61\times10^{2}$ & $1.24\times10^{3}$ &0.939&$8.91\times10^{2}$&$8.26\times10^{2}$\\
$^{171}$Ir$^m\to^{167}$Re$$&6.155&$ {(11/2)-}\to{(9/2-)}$ &2 & $2.70\times10^{0}$ & $1.34\times10^{0}$ &0.174&$3.81\times10^{0}$&$3.40\times10^{0}$\\
$^{173}$Ir$^m\to^{169}$Re$$&5.942&$ {(11/2-)}\to{(9/2-)}$ &2 &$1.82\times10^{1}$&$9.07\times10^{0}$ &0.174&$3.10\times10^{1}$ &$2.91\times10^{1}$\\
$^{175}$Ir$\to^{171}$Re$$&5.431&$ {5/2-\#}\to{(9/2-)}$ &2 & $1.06\times10^{3}$ &$ 1.71\times10^{3}$&0.564&$2.11\times10^{3}$&$2.09\times10^{3}$\\
$^{175}$Pt$\to^{171}$Os$$&6.178&$ {(7/2-)}\to{(5/2-)}$ &2 & $3.95\times10^{0}$ & $2.66\times10^{0}$ &0.235&$5.58\times10^{0}$&$5.40\times10^{0}$\\
$^{177}$Hg$\to^{173}$Pt$$&6.735&$ {(7/2-)}\to{(5/2-)}$ &2 & $1.50\times10^{-1}$ &$1.51\times10^{-1}$&0.353&$8.16\times10^{-2}$&$9.78\times10^{-2}$\\
$^{179}$Pt$\to^{175}$Os$$&5.412&$ {1/2-}\to{(5/2-)}$ &2 & $1.06\times10^{4}$ & $5.75\times10^{3}$&0.190&$1.99\times10^{4}$&$2.13\times10^{4}$\\
$^{179}$Au$\to^{175}$Ir$$&5.980&$ {(1/2+,3/2+)}\to{5/2-\#}$ &1 & $3.23\times10^{1}$& $2.89\times10^{1}$&0.314&$4.00\times10^{1}$&$4.30\times10^{1}$\\
$^{181}$Au$\to^{177}$Ir$$&5.751&$ {(3/2-)}\to{5/2-} $ &2 &$5.07\times10^{2}$ & $4.14\times10^{2}$&0.286&$7.16\times10^{2}$&$8.05\times10^{2}$\\
$^{181}$Hg$\to^{177}$Pt$$&6.284&$ {1/2(-\#)}\to{5/2-}$ &2 & $1.33\times10^{1}$ & $6.51\times10^{0}$&0.171&$1.03\times10^{1}$&$1.27\times10^{1}$\\
$^{181}$Tl$^m\to^{177}$Au$^m$&6.968&${(9/2-)}\to{11/2-}$ &2 & $3.50\times10^{-1}$ &$5.09\times10^{-2}$&0.051&$6.91\times10^{-2}$&$1.60\times10^{-1}$\\

\end{tabular}
\end{ruledtabular}
\end{table*}

\begin{table*}
\caption{Same as Table \ref{table 1}, but for unfavored  $\mathcal{\alpha}$ decay of odd-$A$ nuclei around the doubly $Z=82$, $N=126$ closed shell.}
\label{table 2}
\begin{ruledtabular}
\begin{tabular}{ccccccccc}

{$\mathcal{\alpha}$ transition} & $Q_{\alpha}$ (MeV) &  ${j^{\pi}_{p}}\to{j^{\pi}_{d}}$ &$l_{\text{min}}$ & $T^{\text{expt}}_{1/2}$ (s) & ${T_{1/2}^{\text{calc1}}}$ (s)&${P_{\alpha}}$& ${T_{1/2}^{\text{calc2}}}$  (s)& ${T_{1/2}^{\text{calc3}}}$ (s)\\
\hline
\noalign{\global\arrayrulewidth1pt}\noalign{\global\arrayrulewidth0.4pt} \multicolumn{9}{c}{\textbf{Nuclei in Region III}}\\

$^{187}$Pb$\to^{183}$Hg$$&6.393&$ {3/2-}\to{1/2-}$ &2 &$1.60\times10^{2}$ & $1.54\times10^{1}$&0.034&$2.68\times10^{1}$&$1.84\times10^{1}$\\
$^{189}$Pb$\to^{185}$Hg$$&5.871&$ {3/2-}\to{1/2-}$ & 2&$1.26\times10^{4}$&$2.49\times10^{3}$&0.069& $2.11\times10^{3}$&$1.45\times10^{3}$\\
$^{189}$Po$\to^{185}$Pb$$&7.694&$ {(5/2-)}\to{3/2-}$ &2 &$3.80\times10^{-3}$&$2.62\times10^{-3}$&0.242&$2.49\times10^{-3}$&$2.32\times10^{-3}$\\
$^{191}$At$^m\to^{187}$Bi$$&7.880&${(7/2-)}\to{9/2-\#}$ & 2&$2.20\times10^{-3}$&$1.65\times10^{-3}$& 0.262&$1.90\times10^{-3}$&$1.80\times10^{-3}$\\
$^{193}$Rn$\to^{189}$Po$$&8.040&$ {3/2-\#}\to{(5/2-)}$ & 2&$1.15\times10^{-3}$ &$1.25\times10^{-3}$&0.381&$1.21\times10^{-3}$&$1.14\times10^{-3}$\\
$^{193}$At$^m\to^{189}$Bi$$&7.581&${7/2-\#}\to{(9/2-)}$&2 &$2.10\times10^{-2}$ &$1.27\times10^{-2}$& 0.212&$1.67\times10^{-2}$&$1.68\times10^{-2}$\\
$^{203}$Po$\to^{199}$Pb$$&5.496&$ {5/2-}\to{3/2-}$ & 2&$2.20\times10^{6}$ &$8.97\times10^{5}$&0.143&$7.26\times10^{5}$&$8.19\times10^{5}$\\
$^{205}$Rn$\to^{201}$Po$$&6.390&$ {5/2-}\to{3/2-}$ &2 &$7.08\times10^{2}$ &$4.88\times10^{2}$&0.241&$3.48\times10^{2}$&$4.21\times10^{2}$\\
$^{207}$Ra$\to^{203}$Rn$$&7.274&$ {5/2-\#}\to{3/2-\#}$ & 2&$1.60\times10^{0}$ &$1.30\times10^{0}$& 0.285&$1.05\times10^{0}$&$1.26\times10^{0}$\\
$^{211}$Rn$\to^{207}$Po$$&5.965&$ {1/2-}\to{5/2-}$ & 2&$1.95\times10^{5}$&$2.91\times10^{4}$& 0.052&$4.17\times10^{4}$&$3.77\times10^{4}$\\
$^{213}$Ra$\to^{209}$Rn$$&6.862&$ {1/2-}\to{5/2-}$ &2 &$2.05\times10^{2}$ &$3.29\times10^{1}$&0.056&$4.85\times10^{1}$&$4.63\times10^{1}$\\
$^{215}$Th$\to^{211}$Ra$$&7.665&$ {(1/2-)}\to{5/2(-)}$ &2 & $1.20\times10^{0}$&$2.77\times10^{-1}$&0.081&$3.12\times10^{-1}$&$3.04\times10^{-1}$ \\

\noalign{\global\arrayrulewidth1pt}\noalign{\global\arrayrulewidth0.4pt} \multicolumn{9}{c}{\textbf{Nuclei in Region IV}}\\

$^{209}$Bi$\to^{205}$Tl$$&3.137&$ {9/2-}\to{1/2+}$ & 5&$6.28\times10^{26}$& $5.88\times10^{17}$&$3.28\times10^{-10}$&$ 1.05\times10^{25}$&$1.05\times10^{25}$\\
$^{211}$Po$^m\to^{207}$Pb$^m$&7.423&${(25/2+)}\to{13/2+}$ & 6&$2.52\times10^{1}$&$1.81\times10^{-1}$&0.003&$7.57\times10^{-1}$&$7.75\times10^{-1}$\\
$^{211}$Po$\to^{207}$Pb$$&7.595&$ {9/2+}\to{1/2-}$ &5 &$5.16\times10^{-1}$&$1.64\times10^{-2}$&0.011&$1.59\times10^{-2}$&$1.63\times10^{-2}$\\
$^{213}$Bi$\to^{209}$Tl$$&5.983&$ {9/2-}\to{(1/2+)}$ &5 &$1.31\times10^{5}$&$7.16\times10^{3}$&0.019&$5.88\times10^{3}$&$6.42\times10^{3}$\\
$^{213}$Rn$\to^{209}$Po$$&8.243&$ {9/2+\#}\to{1/2-}$ &5 &$1.95\times10^{-2}$ &$1.04\times10^{-3}$&0.019 &$8.76\times10^{-4}$&$8.75\times10^{-4}$\\
$^{215}$Ra$\to^{211}$Rn$$&8.864&$ {9/2+\#}\to{1/2-}$ & 5&$1.67\times10^{-3}$ & $1.07\times10^{-4}$& 0.023&$9.86\times10^{-5}$&$9.44\times10^{-5}$\\
$^{217}$Th$\to^{213}$Ra$$&9.435&$ {9/2+\#}\to{1/2-}$ & 5&$2.47\times10^{-4}$&$1.85\times10^{-5}$&0.026&$1.81\times10^{-5}$&$1.65\times10^{-5}$\\
$^{219}$Rn$\to^{215}$Po$$&6.946&$ {5/2+}\to{9/2+}$ & 2&$3.96\times10^{0}$&$1.43\times10^{0}$& 0.127&$8.48\times10^{-1}$&$9.37\times10^{-1}$\\
$^{219}$Ra$\to^{215}$Rn$$&8.138&$ {(7/2)+}\to{9/2+}$ &2 &$1.00\times10^{-2}$&$1.06\times10^{-3}$& 0.037&$2.28\times10^{-3}$&$2.32\times10^{-3}$\\
$^{219}$U$\to^{215}$Th$$&9.943&$ {9/2+\#}\to{(1/2-)}$ & 5&$5.50\times10^{-5}$&$5.28\times10^{-6}$&0.034&$4.80\times10^{-6}$&$4.14\times10^{-6}$\\
$^{221}$Rn$\to^{217}$Po$$&6.163&$ {7/2+}\to{(9/2+)}$ & 2&$7.01\times10^{3}$ &$2.13\times10^{3}$&0.106&$1.90\times10^{3}$&$2.17\times10^{3}$\\

\end{tabular}
\end{ruledtabular}
\end{table*}

\begin{table*}
\caption{Same as Table \ref{table 1}, but for unfavored $\mathcal{\alpha}$ decay of doubly-odd nuclei.}
\label{table 3}
\begin{ruledtabular}
\begin{tabular}{ccccccccc}

{$\mathcal{\alpha}$ transition} & $Q_{\alpha}$ (MeV) &  ${j^{\pi}_{p}}\to{j^{\pi}_{d}}$ &$l_{\text{min}}$ & $T^{\text{expt}}_{1/2}$ (s) & ${T_{1/2}^{\text{calc1}}}$ (s)&${P_{\alpha}}$& ${T_{1/2}^{\text{calc2}}}$  (s)& ${T_{1/2}^{\text{calc3}}}$ (s)\\
\hline
 \noalign{\global\arrayrulewidth1pt}\noalign{\global\arrayrulewidth0.4pt} \multicolumn{9}{c}{\textbf{Nuclei in Region I around $Z=82$, $N=82$ shell closure}}\\

$^{154}$Ho$^m\to^{150}$Tb$^m$&3.823&${8+}\to{9+}$ &2 &$1.86\times10^{7}$&$4.23\times10^{7}$&0.410&$2.57\times10^{7}$&$3.34\times10^{7}$\\
$^{168}$Re$\to^{164}$Ta$$&5.068&$ {(7+)}\to{(3+)}$ & 4&$8.80\times10^{4}$&$1.32\times10^{5}$&0.270&$1.27\times10^{5}$&$1.24\times10^{5}$\\
$^{170}$Ir$^m\to^{166}$Re$$&6.265&$ {(8+)}\to{(7+)}$ & 2&$ 2.25\times10^{0}$&$1.01\times10^{0}$&0.081&$1.86\times10^{0}$&$1.36\times10^{0}$\\
$^{172}$Ir$\to^{168}$Re$$&5.985&$ {(3+)}\to{(7+)}$ &4 &$2.20\times10^{2}$&$5.42\times10^{1}$&0.044&$2.43\times10^{2}$&$2.26\times10^{2}$\\
$^{180}$Tl$\to^{176}$Au$$&6.715&$ {4(-)}\to{(5-)}$ &2 &$1.70\times10^{1}$&$8.47\times10^{-1}$&0.009&$-3.92\times10^{0}$ &$-5.18\times10^{0}$\\

 \noalign{\global\arrayrulewidth1pt}\noalign{\global\arrayrulewidth0.4pt} \multicolumn{9}{c}{\textbf{Nuclei in Region III around $Z=82$, $N=126$ shell closure}}\\

$^{190}$Bi$\to^{186}$Tl$^m$&6.836&$  {(3+)}\to{(7+)}$  & 4&$8.16\times10^{0}$&$5.72\times10^{0}$ & 0.126&$1.00\times10^{1}$&$8.70\times10^{0}$\\
$^{192}$Bi$\to^{188}$Tl$^m$&6.343&$  {(3+)}\to{(7+)}$  & 4&$2.77\times10^{2}$&$4.43\times10^{2}$&0.288&$3.00\times10^{2}$&$2.80\times10^{2}$\\
$^{192}$Bi$^m\to^{188}$Tl$^n$&6.207&${(10-)}\to{(9-)}$  &2 &$3.84\times10^{2}$&$4.12\times10^{2}$&0.193&$4.17\times10^{2}$&$3.89\times10^{2}$\\
$^{194}$Bi$^n\to^{190}$Tl$^p$&5.696&${(10-)}\to{9-}$  &2 &$5.56\times10^{4}$ &$8.18\times10^{4}$&0.265&$5.22\times10^{4}$&$5.21\times10^{4}$\\
$^{210}$At$\to^{206}$Bi$$&5.631&$    {(5)+}\to{6(+)}$  & 2&$1.66\times10^{7}$&$8.14\times10^{5}$&0.009& $1.07\times10^{6}$&$1.18\times10^{6}$\\
$^{210}$Fr$\to^{206}$At$$&6.671&$    {6+}\to{(5)+}$  &2 &$2.67\times10^{2}$ &$1.48\times10^{2}$&0.100&$2.51\times10^{2}$&$3.09\times10^{2}$\\
$^{212}$Fr$\to^{208}$At$$&6.529&$    {5+}\to{6+}$  &2 &$2.78\times10^{3}$ &$5.10\times10^{2}$& 0.033&$5.84\times10^{2}$&$6.88\times10^{2}$\\
$^{214}$Ac$\to^{210}$Fr$$&7.353&$    {5+\#}\to{6+}$  &2 &$9.18\times10^{0}$&$2.51\times10^{0}$&0.049&$3.27\times10^{0}$&$3.68\times10^{0}$\\

 \noalign{\global\arrayrulewidth1pt}\noalign{\global\arrayrulewidth0.4pt} \multicolumn{9}{c}{\textbf{Nuclei in Region IV around $Z=82$, $N=126$ shell closure}}\\

$^{212}$Bi$^m\to^{208}$Tl$$&6.455&${(8-,9-)}\to{5+}$ &3 &$2.24\times10^{3}$ &$2.40\times10^{1}$&0.002&$5.64\times10^{1}$&$6.83\times10^{1}$\\
$^{212}$At$\to^{208}$Bi$$&7.817&$ {(1-)}\to{5+}$ &5 &$3.14\times10^{-1}$&$1.57\times10^{-2}$&0.009&$7.92\times10^{-3}$&$8.20\times10^{-3}$\\
$^{214}$Bi$\to^{210}$Tl$$&5.621&$ {-1}\to{5+\#}$ &5 &$5.66\times10^{6}$&$7.18\times10^{5}$&0.023&$3.04\times10^{5}$&$3.80\times10^{5}$\\
$^{214}$Fr$^m\to^{210}$At$$&8.710&$ {(8-)}\to{(5)+}$ &3 &$3.35\times10^{-3}$&$4.60\times10^{-5}$&0.002&$1.80\times10^{-4}$&$1.76\times10^{-4}$\\
$^{216}$Ac$\to^{212}$Fr$$&9.236&$ {(1-)}\to{5+}$ &5 &$4.40\times10^{-4}$&$5.11\times10^{-5}$ &0.021&$3.62\times10^{-5}$&$3.33\times10^{-5}$\\
$^{216}$Ac$^m\to^{212}$Fr$$&9.279&${(9-)}\to{5+}$ & 5&$4.41\times10^{-4}$&$4.00\times10^{-5}$ & 0.016&$3.63\times10^{-5}$&$3.34\times10^{-5}$\\
$^{220}$Fr$\to^{216}$At$$&6.800&$ {1+}\to{1(-)}$ &1 &$2.74\times10^{1}$&$1.86\times10^{1}$&0.122&$1.32\times10^{1}$&$1.51\times10^{1}$\\
$^{224}$Pa$\to^{220}$Ac$$&7.694&$ {5-\#}\to{(3-)}$ &2 &$8.44\times10^{-1}$&$7.15\times10^{-1}$&0.152 &$7.44\times10^{-1}$&$7.60\times10^{-1}$\\

\end{tabular}
\end{ruledtabular}
\end{table*}

In Region I, proton numbers are below $Z=82$ shell closure, and neutron numbers above $N=82$ closed shell. Therefore, the $N_p N_n$ are negative. The calculations of odd-$A$ nuclei in Region I are listed in Table \ref{table 1}. In Region III, the proton numbers are above the $Z=82$ closed shell, but the neutron numbers below the $N=126$ closed shell, so the $N_p N_n$ are negative. Similarly, in the Region IV the $N_p N_n$ are positive. The detailed calculations of odd-$A$ nuclei in Region III, IV are given in upper half and bottom half of Table \ref{table 2}, respectively.

\begin{figure}[b]
\includegraphics[width=8.5cm]{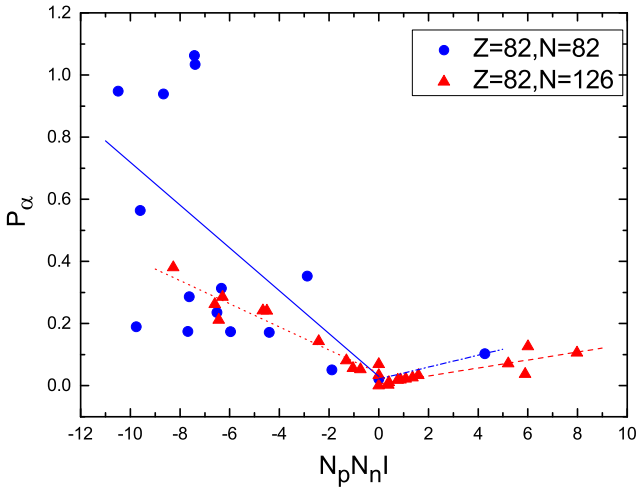}
\caption{(color online) Same as Fig. \ref{fig 2}, but it depicts linear relationship between $\mathcal{\alpha}$ preformation probabilities and product of valence protons (holes), neutrons (holes) and isospin asymmetry as $N_p N_nI$.}
\label{fig 3}
\end{figure}

\begin{figure}[b]
\includegraphics[width=8.5cm]{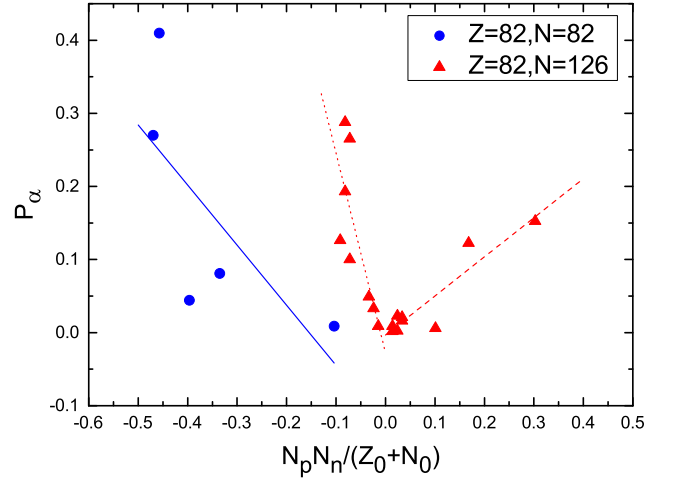}
\caption{(color online) Same as Fig. \ref{fig 2}, but it depicts doubly-odd nuclei in accordance with $\frac{N_p N_n}{Z_0+N_0}$.}
\label{fig 4}
\end{figure}

\begin{figure}[b]
\includegraphics[width=8.5cm]{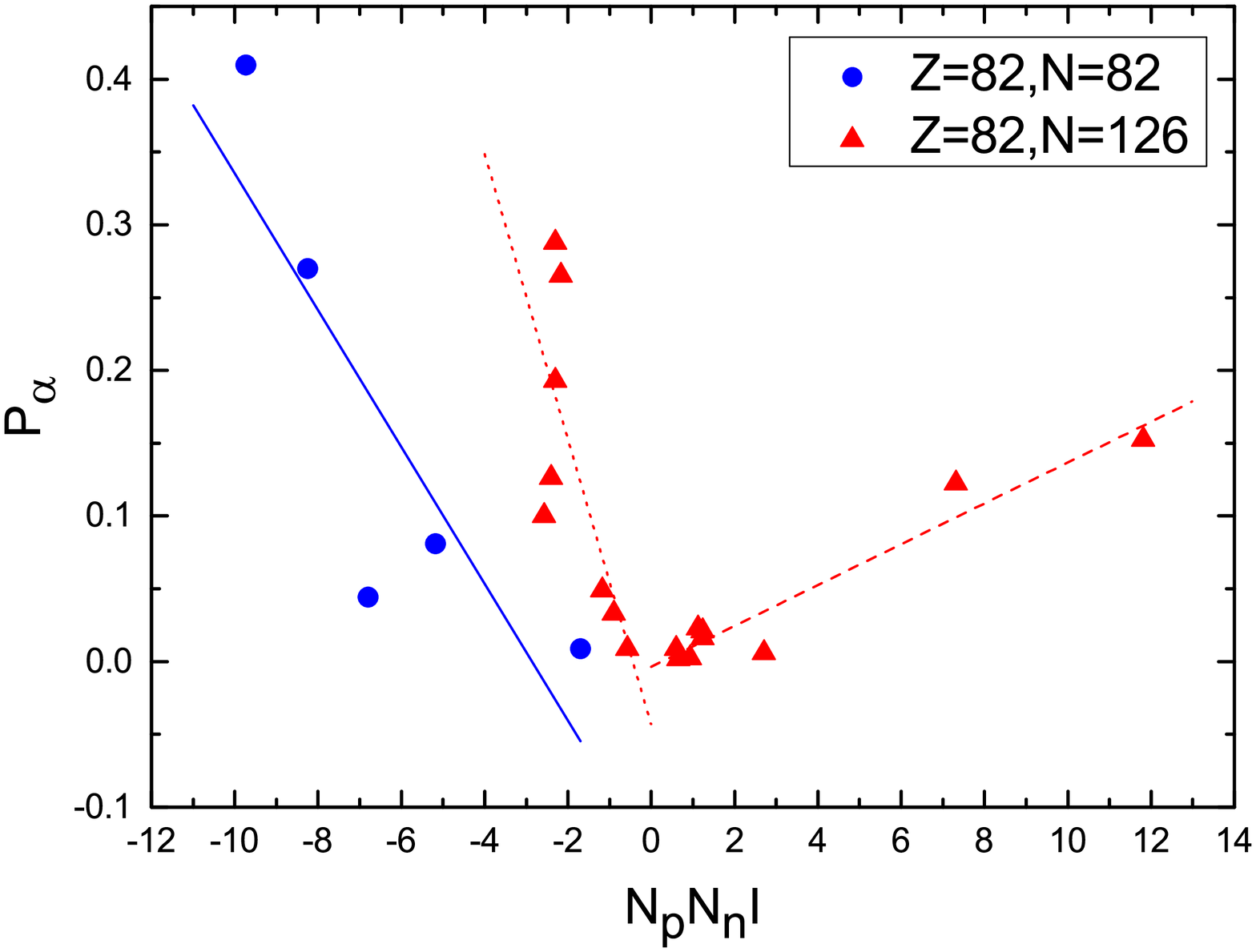}
\caption{(color online) Same as Fig. \ref{fig 2}, but it depicts doubly-odd nuclei in accordance with $N_p N_n I$.}
\label{fig 5}
\end{figure}

For the unfavored $\mathcal{\alpha}$ decay of doubly-odd nuclei, the detailed calculations are listed in Table \ref{table 3}, in this table, Regions I, III, and IV are involved.  From Table \ref{table 1}-\ref{table 3}, we can find that  last two columns ${T_{1/2}^{\text{calc2}}}$, ${T_{1/2}^{\text{calc3}}}$ are well conform because of Eq. (\ref{subeq:11}) and (\ref{subeq:12}) are two different perspectives for studying the liner relationships between $N_pN_n$ and $P_{\alpha}$. From Table \ref{table 4}, we can clearly see that the values of adjustable parameters $b$ and $d$ in Eq. (\ref{subeq:11}) and (\ref{subeq:12}) are approximatively equal in each region because the isospin $I$ changes little.

For intuitively, the linear relationships of unfavored $\mathcal{\alpha}$ decay for odd-$A$ nuclei between $P_{\alpha}$ and $\frac{N_p N_n}{Z_0+N_0}$, ${ N_p N_n I}$ as Eq. (\ref{subeq:11}), Eq. (\ref{subeq:12}) are plotted in Fig. \ref{fig 2} and \ref{fig 3}, respectively. Similarly, for the unfavored $\mathcal{\alpha}$ decay of doubly-odd nuclei, the corresponding linear relationships are depicted in Fig. \ref{fig 4} and \ref{fig 5}, respectively.  In Fig. \ref{fig 2}-\ref{fig 5}, the blue circle and the red triangle represent the nuclei around at $Z=82$, $N=82$ and $Z=82$, $N=126$ closed shells, respectively. The lines are linear fittings between $P_{\alpha}$ and $\frac{N_p N_n}{Z_0+N_0}$, ${N_p N_n I}$, respectively, and also are predictions of Eq. (\ref{subeq:11}) and (\ref{subeq:12}). The corresponding parameters $a$, $b$, $c$, and $d$ are given in the Table \ref{table 4}. From Fig. \ref{fig 2} and \ref{fig 3}, we can find that valence proton-neutron interaction have an obvious difference in different shell closures. From Fig. \ref{fig 4} and \ref{fig 5}, we can intuitively find that the lines' variation tendencies still satisfy above equations although the nuclei number of unfavored $\mathcal{\alpha}$ decay is small.

Intuitively, we can find that the linear relationship in Regions III and IV are better than those in Region I from Fig. \ref{fig 2}-\ref{fig 5}. It might because the doubly magic core at $Z=82$, $N=82$ is unbound, and the nucleons in the core play an essential role on $P_{\alpha}$ \cite{PhysRevC.94.024338}. Meanwhile, $Z=82$, $N=126$ are stable doubly magic core.

\begin{table}[!htb]
\caption{The parameters of Eqs. (\ref{subeq:11}) and (\ref{subeq:12})that show $\mathcal{\alpha}$ preformation probabilities are linearly related to $N_pN_n$.}
\label{table 4}
\begin{ruledtabular}
\begin{tabular}{cccccccc}

Region&a&b&c&d\\
\hline
\noalign{\global\arrayrulewidth1pt}\noalign{\global\arrayrulewidth0.4pt} \multicolumn{5}{c}{\textbf{odd-$A$ Nuclei}}\\
I&-1.65948&-0.11308&-0.06898&0.02948\\
III&-0.8437&0.05854&-0.03726&0.0402\\
IV&0.51361&0.00585&0.01281&0.00585\\
\noalign{\global\arrayrulewidth1pt}\noalign{\global\arrayrulewidth0.4pt} \multicolumn{5}{c}{\textbf{doubly-odd Nuclei}}\\
I&-0.82097&-0.12653&-0.04695&-0.13455\\
III&--2.72853&-0.02778&-0.09794&-0.04321\\
IV&0.53443&-0.00317&0.01402&-0.00363\\

\end{tabular}
\end{ruledtabular}
\end{table}

\section{Summary}
\label{section 4}
In summary, we systematically study unfavored $\mathcal{\alpha}$ decay of closed shell nuclei related to ground and isomeric states around $Z=82$, $N=82$ and 126 closed shells, respectively, within a two-potential approach. Our research indicates that, for unfavored $\alpha$ decay of closed shell nuclei, $P_{\alpha}$ are still linear to $N_p N_n$ or $N_p N_n I$, in addition, shell effect and valence proton-neutron interaction still plays an important role in $P_{\alpha}$. Our calculations are in good agreement with the experimental data.
\begin{acknowledgments}

We would like to thank Xiao-Dong Sun for useful discussions. This work is supported in part by the National Natural Science Foundation of China (Grants No. 11205083 and No. 11505100), the Construct Program of the Key Discipline in Hunan Province, the Research Foundation of Education Bureau of Hunan Province, China (Grant No. 15A159), the Natural Science Foundation of Hunan Province, China (Grants No. 2015JJ3103 and No. 2015JJ2121), the Innovation Group of Nuclear and Particle Physics in USC, the Shandong Province Natural Science Foundation, China (Grant No. ZR2015AQ007), Hunan Provincial Innovation Foundation For Postgraduate (Grant No. CX2017B536).

\end{acknowledgments}


%

\end{document}